\documentclass[article,preprint,groupedaddress,amsmath,amssymbm]{revtex4}
\usepackage{epsfig}
\usepackage{graphicx}

\begin{document}
\title{Quantum-Gravity Fluctuations and the Black-Hole Temperature}
\author{Shahar Hod}
\address{The Ruppin Academic Center, Emeq Hefer 40250, Israel}
\address{ }
\address{The Hadassah Institute, Jerusalem 91010, Israel}
\date{\today}

\begin{abstract}

\ \ \ Bekenstein has put forward the idea that, in a quantum theory
of gravity, a black hole should have a discrete energy spectrum with
concomitant discrete line emission. The quantized  black-hole
radiation spectrum is expected to be very different from Hawking's
semi-classical prediction of a thermal black-hole radiation
spectrum. One naturally wonders: Is it possible to reconcile the
{\it discrete} quantum spectrum suggested by Bekenstein with the
{\it continuous} semi-classical spectrum suggested by Hawking ?
\newline
In order to address this fundamental question, in this essay we
shall consider the zero-point quantum-gravity fluctuations of the
black-hole spacetime. In a quantum theory of gravity, these
spacetime fluctuations are closely related to the characteristic
gravitational resonances of the corresponding black-hole spacetime.
Assuming that the energy of the black-hole radiation stems from
these zero-point quantum-gravity fluctuations of the black-hole
spacetime, we derive the effective temperature of the quantized
black-hole radiation spectrum. Remarkably, it is shown that this
characteristic temperature of the {\it discrete} (quantized)
black-hole radiation agrees with the well-known Hawking temperature
of the {\it continuous} (semi-classical) black-hole spectrum.
\newline
\newline

\end{abstract}
\bigskip
\maketitle

%]

One of the most remarkable theoretical predictions of modern physics
is Hawking's celebrated result that black holes are not completely
black \cite{Haw1}. According to Hawking's semi-classical analysis, a
black hole is quantum mechanically unstable -- it emits continuous
thermal radiation whose characteristic temperature is given by
\begin{equation}\label{Eq1}
T_{\text{H}}={{\hbar}\over{8\pi M}}\  .
\end{equation}
Here $M$ is the mass of the Schwarzschild black hole. (We use
gravitational units in which $G=c=1$.)

It should be stressed, however, that Hawking's derivation of the
continuous  black-hole radiation spectrum is restricted to the
semi-classical regime: the matter fields are treated quantum
mechanically but the spacetime (and, in particular, the black hole
itself) are treated classically. One therefore expects to find
important new features in the character of the black-hole radiation
spectrum once quantum properties of the {\it black hole} itself are
properly taken into account \cite{Notequant}. It is therefore
appropriate to regard the Hawking temperature (\ref{Eq1}) as the
{\it semi-classical} (SC) temperature of the continuous black-hole
radiation:
\begin{equation}\label{Eq2}
T_{\text{SC}}\equiv T_{\text{H}}={{\hbar}\over{8\pi M}}\  .
\end{equation}

The quantization of black holes was first proposed in the seminal
work of Bekenstein \cite{Beken1,Beken2}. The original quantization
procedure was based on the physical observation that the surface
area of a black hole behaves as a classical adiabatic invariant
\cite{Beken1,Beken2}. In the spirit of Ehrenfest principle
\cite{Ehr}, any classical adiabatic invariant corresponds to a
quantum entity with a discrete spectrum, Bekenstein suggested that
the horizon area $A$ of a quantum black hole should have a discrete
spectrum of the form
\begin{equation}\label{Eq3}
A_n=\gamma\hbar\cdot n\ \ \ ;\ \ \ n=1,2,3,\ldots\ \  .
\end{equation}
Here $\gamma$ is an unknown ``fudge" factor which was introduced in
\cite{Beken1,Beken2}.

In order to determine the value of the coefficient $\gamma$,
Mukhanov and Bekenstein \cite{Muk,BekMuk} have suggested, in the
spirit of the Boltzmann-Einstein formula in statistical physics
\cite{Ehr}, to relate $g_n \equiv \exp[S_{\text{BH}}(n)]$ to the
number of black-hole micro-states that correspond to a particular
external black-hole macro-state. Here $S_{\text{BH}}$ is the
black-hole entropy, which is related to its surface area $A$ by the
thermodynamic-geometric relation \cite{Haw1,Beken1}
\begin{equation}\label{Eq4}
S_{\text{BH}}={{A}\over{4\hbar}}\  .
\end{equation}
The statistical degeneracy [see Eqs. (\ref{Eq3}) and (\ref{Eq4})]
\begin{equation}\label{Eq5}
g_n\equiv\exp[S_{\text{BH}}(n)]=\exp({1\over4}\gamma\cdot n)
\end{equation}
of the $n$th black-hole area level has to be an integer for every
integer $n$. This physical requirement dictates the relation
\cite{Muk,BekMuk}
\begin{equation}\label{Eq6}
\gamma =4\ln{k}\
\end{equation}
for the fudge factor $\gamma$, where the unknown constant $k$ should
be an integer.

Determining the specific value of the integer $k$ requires
additional physical input. This physical information emerges by
applying the Bohr correspondence principle \cite{Ehr} to the
discrete resonance spectrum of the black hole \cite{Hod2}. According
to the Bohr correspondence principle, transition frequencies at
large quantum numbers should equal classical oscillation
frequencies. Namely, the asymptotic energy difference $M_{n+1}-M_n$
between the $(n+1)$th and the $n$th black-hole quantum levels should
be given by the characteristic classical oscillation frequency of
the black hole:
\begin{equation}\label{Eq7}
M_{n+1}-M_n=\hbar\omega_{\text{BH}}\  .
\end{equation}

It is well-known that a Schwarzschild black hole is characterized by
a discrete spectrum of gravitational resonances
\cite{Nollert1,Leaver,Nollert2} with the fundamental asymptotic
frequency \cite{Hod2,Noteasym}
\begin{equation}\label{Eq8}
M\omega_{\text{R}}={{\ln3}\over{8\pi}}\  .
%\omega_{\text{R}}=T_{\text{BH}}\ln3\  .
\end{equation}
The emission of a gravitational quantum from the black hole results
in a change $\Delta M=\hbar \omega_R$ [see Eq. (\ref{Eq7})] in the
black-hole mass. Using the first-law of black-hole thermodynamics,
$\Delta A=32\pi M\Delta M$ \cite{Noteam}, one finds the fundamental
change $\Delta A =4\ln3\cdot\hbar$ in the Schwarzschild black-hole
surface area. Taking cognizance of Eqs. (\ref{Eq3}) and (\ref{Eq8}),
one finally obtains the quantized area spectrum:
\begin{equation}\label{Eq9}
A_n=4\hbar\ln3\cdot n\ \ \ ;\ \ \ n=1,2,3,\ldots\ \  .
\end{equation}
It is worth emphasizing again that the black-hole area spectrum
(\ref{Eq9}) is consistent both with the area-entropy thermodynamic
relation (\ref{Eq4}) for black holes, with the Boltzmann-Einstein
formula (\ref{Eq5}) in statistical physics, and with the Bohr
correspondence principle (\ref{Eq7}) \cite{Hod2}.

One therefore concludes that, in a quantum theory of gravity, a
Schwarzschild black hole has a discrete energy (mass) spectrum of
the form \cite{Noteam}:
\begin{equation}\label{Eq10}
M_n=\sqrt{{{\hbar\ln3}\over{4\pi}}}\cdot n^{1/2}\ \ \ ;\ \ \
n=1,2,3,\ldots\ \  ,
\end{equation}
with concomitant {\it discrete} line emission
\cite{Muk,BekMuk,Hod2}. In particular, the radiation emitted by the
quantized black hole consists of gravitational quanta whose
frequencies are integer multiples of the fundamental black-hole
frequency
\begin{equation}\label{Eq11}
\omega_0\equiv (M_{n+1}-M_n)/\hbar={{\ln3}\over{8\pi M}}\  .
\end{equation}
The quantized (discrete) black-hole radiation spectrum is obviously
different from Hawking's semi-classical prediction of a thermal
(continuous) spectrum.

One naturally wonders: Is it possible to reconcile the {\it
discrete} quantum spectrum predicted by Bekenstein with the {\it
continuous} semi-classical spectrum predicted by Hawking ?
\newline
In order to address this fundamental question, we first point out
that, in a quantum theory of gravity, the black-hole spacetime is
expected to possess a set of zero-point quantum-gravity
fluctuations. It has been suggested \cite{York,HarHaw} that these
zero-point fluctuations of the black-hole spacetime (and, in
particular, the quantum-gravity fluctuations of the black-hole
horizon) may enable quanta to tunnel out of the black hole.

We shall now conjecture that these black-hole spacetime fluctuations
are characterized by the fundamental resonance frequency $\omega_0$
[see Eq. (\ref{Eq11})] of the black-hole spacetime. In particular,
we propose a physical picture in which the quantum-gravity
fluctuations of the black-hole spacetime may enable quanta with the
appropriate frequencies (the ones which are in resonance with the
fluctuating horizon: $\omega_0,2\omega_0,3\omega_0,...$) to tunnel
out of the quantum black hole.

According to this physical picture, the energy of the black-hole
radiation stems from these zero-point quantum-gravity fluctuations
of the black-hole spacetime. The characteristic quantum temperature,
$T_{\text{Q}}$, of the discrete black-hole radiation spectrum may be
defined by equating the mean thermal energy of the radiating fields
(including their zero-point quantum energy) with the corresponding
energy of an emitted quantum with the characteristic black-hole
resonance frequency $\omega_0$. Namely,
\begin{equation}\label{Eq12}
{1 \over 2}\hbar \omega_0+{{\hbar \omega_0} \over {e^{\hbar\omega_0
/T_{\text{Q}}}-1}}=\hbar \omega_0\  .
\end{equation}
Substituting the fundamental black-hole resonance
$\omega_0=\ln3/8\pi M$ [see Eq. (\ref{Eq11})] into equation
(\ref{Eq12}), one finds that the characteristic temperature of the
{\it quantized} (Q) black-hole radiation spectrum is given by
\begin{equation}\label{Eq13}
T_{\text{Q}}={{\hbar}\over{8\pi M}}\  .
\end{equation}

Remarkably, we find here that the characteristic temperature
(\ref{Eq13}) of the {\it discrete} (quantized) black-hole radiation
spectrum exactly matches the semi-classical temperature (\ref{Eq2})
of the {\it continuous} black-hole radiation:
\begin{equation}\label{Eq14}
T_{\text{Q}}=T_{\text{SC}}\  .
\end{equation}

{\it Summary}.\ --- One of the most important theoretical
predictions of modern physics is Hawking's semi-classical
\cite{Notesemi} result that black holes are not completely black
\cite{Haw1}. In particular, according to Hawking's analysis, a
Schwarzschild black hole is expected to emit {\it continuous}
thermal radiation whose characteristic semi-classical temperature is
given by Eq. (\ref{Eq2}). However, Bekenstein \cite{Beken1,Beken2}
has put forward the idea that, in a quantum theory of gravity
\cite{Notequan}, a quantum black hole should have a discrete mass
spectrum. As a consequence, a quantum black hole is expected to be
characterized by a {\it discrete} line emission.

In the present essay we have proposed a physical mechanism which
relates in a natural way the two seemingly different spectra: the
{\it discrete} black-hole quantum spectrum as predicted by
Bekenstein and the semi-classical {\it continuous} spectrum as
predicted by Hawking. The proposed model is based on the fact that,
in a quantum theory of gravity, the black-hole spacetime is expected
to possess a set of zero-point quantum-gravity fluctuations which
are characterized by the fundamental black-hole resonance frequency
(\ref{Eq11}). These quantum-gravity fluctuations of the black-hole
horizon may enable quanta with the appropriate frequencies (the
frequencies $\omega_0,2\omega_0,3\omega_0,...$ which are in
resonance with the fluctuating horizon) to tunnel out of the quantum
black hole.

The resulting quantum-gravity black-hole radiation spectrum is
characterized by a {\it discrete} line emission as predicted by
Bekenstein. Remarkably, we have shown here that this quantized
(discrete) black-hole radiation spectrum is characterized by an
effective quantum temperature $T_{\text{Q}}$ [see Eq. (\ref{Eq13})]
which {\it agrees} with the well-known semi-classical Hawking
temperature $T_{\text{H}}$ [see Eq. (\ref{Eq2})] of the continuous
black-hole radiation spectrum.

\bigskip
\noindent
{\bf ACKNOWLEDGMENTS}
\bigskip

This research is supported by the Carmel Science Foundation. I thank
Yael Oren, Arbel M. Ongo, Ayelet B. Lata, and Alona B. Tea for
stimulating discussions.


\newpage

\begin{thebibliography}{99}

\bibitem{Haw1} S. W. Hawking, Commun. Math. Phys. {\bf 43}, 199 (1975).

\bibitem{Notequant} This state of affairs is reminiscent of atomic spectroscopy:
according to the classical laws of electrodynamics an atom should
have a {\it continuous} emission spectrum, whereas quantum mechanics
dictates a {\it discrete} line emission from the atom.

\bibitem{Beken1} J. D. Bekenstein, Phys. Rev. D {\bf 7}, 2333 (1973).

\bibitem{Beken2} J. D. Bekenstein, Lett. Nuovo Cimento {\bf 11}, 467 (1974).

\bibitem{Ehr} M. Born, {\it Atomic Physics} (Blackie, London, 1969).

\bibitem{Muk} V. Mukhanov, JETP Lett. {\bf 44}, 63 (1986).

\bibitem{BekMuk} J. D. Bekenstein and V. F. Mukhanov, Phys. Lett. B {\bf 360}, 7 (1995).

\bibitem{Hod2} S. Hod, Phys. Rev. Lett. {\bf 81}, 4293 (1998).

\bibitem{Nollert1} H. P. Nollert, Class. Quantum Grav. {\bf 16}, R159 (1999).

\bibitem{Leaver} E. W. Leaver, Proc. R. Soc. A {\bf 402}, 285 (1985).

\bibitem{Nollert2} H. P. Nollert, Phys. Rev. D {\bf 47}, 5253 (1993).

\bibitem{Noteasym} See S. Hod, Class. Quant. Grav. {\bf 23}, L23 (2006) for
the physical motivation of focusing on the asymptotic black-hole
resonances.

\bibitem{Noteam} Here we have used the relation $A=16\pi M^2$ for
the Schwarzschild black hole.

\bibitem{York} J. W. York, Phys. Rev. D {\bf 28}, 2929 (1983).

\bibitem{HarHaw} J. B. Hartle and S. W. Hawking, Phys. Rev. D {\bf
13}, 2188 (1975).

\bibitem{Notesemi} It is worth emphasizing again that,
in Hawking's original analysis \cite{Haw1}, the matter fields are
treated quantum mechanically but the black hole itself is treated as
a classical entity. Thus, Hawking's analysis is restricted to the
semi-classical regime.

\bibitem{Notequan} In a quantum theory of gravity, the spacetime itself (and, in particular,
the black-hole energy spectrum) should be treated as a quantum
entity.

\end{thebibliography}
\end{document}